\documentclass[epj]{svjour}
\usepackage[german,english]{babel}
\usepackage{times}
\usepackage{epsfig}
\usepackage[centertags]{amsmath}
\usepackage{./styles/upgreek}
\usepackage{amssymb,amscd}
\usepackage{cite}
\usepackage{./styles/mcite}
\usepackage{./styles/myxspace}
\def\zero{{\scriptscriptstyle 0}}

\def\Euro{\hbox{\kern0.15em C\kern-0.7em%
\raisebox{-0.1ex}{--}\kern-0.43em\raisebox{0.15ex}{--}\kern0.13em}}

\def\Z0{{Z^\zero}}
\def\eVdist{\kern-0.06667em}

\def\gev{{\,\text{Ge}\eVdist\text{V\/}}}
\def\tev{{\,\text{Te}\eVdist\text{V\/}}}

\def\met{\,\text{m}}

\def\cm{\,\text{cm}}

\def\km{\,\text{km}}

\def\nm{\,\text{nm}}

\def\kHz{\,\text{kHz}}

\def\Scnd{\text{s}}

\def\ns{\,\text{ns}}

\def\IP{{\rm I$\kern-0.01667em$P}\xspace}

\def\Ptlj{{\not{\kern-0.55ex P}}_t\ell j}
\def\Ptmiss{{\not{\kern-0.55ex P}}_t}

\def\rnge{{\,\text{--}\,}}

\newcommand{\nucl}[2]{\ensuremath{\stru{2.ex}{0.8ex}^{#2}{\rm #1}}}

\mathchardef\qsm=63
\mathchardef\pls=43
\mathchardef\mns=512
\mathchardef\plm=518
\mathchardef\eql=61
\mathchardef\smallleft=300
\mathchardef\smallright=301
\mathchardef\perslsh=47
\mathchardef\les=316
\mathchardef\gre=318
\mathchardef\leq=532
\mathchardef\grq=533
\chardef\usc=95
\chardef\til=126

\def\sqr#1#2#3{{\vcenter{\hrule height.#3ex\hbox{\vrule width.#2ex height#1ex
    \kern#1ex\vrule width.#3ex}\hrule height.#2ex}}}

\def\angleto{\vrule width.035em height2.1ex depth-.56ex\unskip\kern-.6ex\to}
\def\perchc#1{{\raise.4ex\hbox{$\mkern4mu#1{\it\perslsh}_
             {\mkern-5mu\scriptscriptstyle{{\rm o}\!{\rm o}}}^
             {\mkern-12.8mu\scriptscriptstyle{\rm o}}$}}}

\catcode`\@=11 
\def\parenbar{\mathpalette\p@renb@r}
\def\p@renb@r#1#2{\vbox{%
  \ifx#1\scriptscriptstyle \dimen@.7em\dimen@ii.2em\else
  \ifx#1\scriptstyle \dimen@.8em\dimen@ii.25em\else
  \dimen@1em\dimen@ii.4em\fi\fi \offinterlineskip
  \ialign{\hfill##\hfill\cr
    \vbox{\hrule width\dimen@ii}\cr
    \noalign{\vskip-.3ex}%
    \hbox to\dimen@{$\mathchar300\hfil\mathchar301$}\cr
    \noalign{\vskip-.3ex}%
    $#1#2$\cr}}}
\catcode`\@=12 

\newbox\struttbox
\setbox\struttbox=\hbox{\vrule height1.65ex depth.485ex width0pt}
\def\strutt{\relax\ifmmode\copy\struttbox\else\unhcopy\struttbox\fi}
\def\stru#1#2{\relax\ifmmode\hbox{\vrule height#1 depth#2 width0pt}
\else\vrule height#1 depth#2 width0pt\fi}
\def\uline#1{$\underline{\hbox{#1\strutt}}$}
\def\ronum#1{\uppercase\expandafter{\romannumeral#1}}
\def\ronuml#1{\expandafter{\romannumeral#1}}

\def\cbk{\kern-0.5em}

\newcommand{\linebox}[2][3.ex]{\uline{\hbox to #2{\stru{#1}{0.pt}\hfil}}}

\DeclareMathAlphabet{\mathbf}{OT1}{cmr}{bx}{n}

\DeclareMathAlphabet{\mathbfs}{OT1}{lcmss}{bx}{sl}

\newlength\listtextwidth

\catcode`\@=11 
\newlength{\@tabfninsert}
\newlength{\@tabfnwidth}
\newcommand{\tabfootnote}[2]{%
  \setlength{\@tabfninsert}{0.8em}
  \setlength{\@tabfnwidth}{\textwidth}
  \addtolength{\@tabfnwidth}{-\@tabfninsert}
  \addtolength{\@tabfnwidth}{-0.4em}
  \noindent\makebox[\@tabfninsert][r]{\footnotesize$^{#1}$\hfil}\hfill%
  \parbox[t]{\@tabfnwidth}{\footnotesize #2\hfill}}
\catcode`\@=12 
\newcommand{\boldarrayrulewidth}{1pt}

\catcode`\@=11 
\let\tab@penalty\relax
\newcount\tab@state
\def\bcline#1{%
  \noalign{\kern-.5\arrayrulewidth\tab@penalty}%
  \omit%
  \global\tab@state\@ne%
  \ranges\bcline@i{#1}%
  \cr%
  \noalign{\kern-.5\arrayrulewidth\tab@penalty}%
}
\def\bcline@i#1#2{%
  \ifnum#1<\tab@state\relax%
    \tab@@cr%
    \noalign{\kern-\arrayrulewidth\tab@penalty}%
    \omit%
    \global\tab@state\@ne%
  \fi%
  \@whilenum\tab@state<#1\do{%
    \hfil\tab@@tab@omit%
    \global\advance\tab@state\@ne%
  }%
  \ifnum\tab@state>\@ne%
    \kern-\arrayrulewidth%
  \fi%
  \@whilenum\tab@state<#2\do{%
    \tab@@span@omit%
    \global\advance\tab@state\@ne%
  }%
  \leaders\hrule\@height\boldarrayrulewidth\hfill%
}
\def\ranges#1#2{%
  \gdef\ranges@temp{#1}%
  \begingroup%
  \ranges@i#2 \q@delim%
}
\def\ranges@i{%
  \@ifnextchar\q@delim\ranges@done{\afterassignment\ranges@ii\count@}%
}
\def\ranges@ii{%
  \@ifnextchar-\ranges@iii{\ranges@do\count@\count@\ranges@v}%
}
\def\ranges@iii-{\afterassignment\ranges@iv\@tempcnta}
\def\ranges@iv{\ranges@do\count@\@tempcnta\ranges@v}
\def\ranges@v{%
  \@ifnextchar,%
    \ranges@vi%
    {%
      \@ifnextchar\q@delim%
        \ranges@done%
        {\tab@err@range\ranges@vi,}%
    }%
}
\def\ranges@vi,{\afterassignment\ranges@ii\count@}
\def\ranges@do#1#2{%
  \ifnum#1>#2\else%
    \expandafter\endgroup%
    \expandafter\ranges@temp%
    \expandafter{%
    \the\expandafter#1%
    \expandafter}%
    \expandafter{%
    \the#2%
    }%
    \begingroup%
  \fi%
}
\def\ranges@done\q@delim{\endgroup}
\def\ifinrange#1#2{%
  \@tempswafalse%
  \count@#1%
  \ranges\ifinrange@i{#2}%
  \if@tempswa%
    \expandafter\@firstoftwo%
  \else%
    \expandafter\@secondoftwo%
  \fi%
}
\def\ifinrange@i#1#2{%
  \ifnum\count@<#1 \else\ifnum\count@>#2 \else\@tempswatrue\fi\fi%
}
\def\tab@@cr{\cr}
\def\tab@@tab@omit{&\omit}
\def\tab@@span@omit{\span\omit}
\def\tab@checkrule#1{%
  \count@#1\relax%
  \expandafter\ifinrange%
  \expandafter\count@%
  \expandafter{\tab@xcols}%
    {\tab@checkrule@i}%
    {}%
}
\def\bhline{\noalign{\ifnum0=`}\fi\hrule \@height  
\boldarrayrulewidth \futurelet \@tempa\@xhline}
\def\@xhline{\ifx\@tempa\hline\vskip \doublerulesep\fi
      \ifnum0=`{\fi}}
\catcode`\@=12 

\catcode`\@=11 
\newcounter{pict@width}
\newcounter{pict@height}
\newlength{\pict@scale}
\newcommand{\psfigadd}[4]{%
\setcounter{pict@width}{1*\ratio{#2+\pict@scale/2}{\pict@scale}}
\setcounter{pict@height}{1*\ratio{#3+\pict@scale/2}{\pict@scale}}
\setlength{\unitlength}{\pict@scale}
\hbox to #2{\hspace{-\fill}\begin{picture}(\thepict@width,\thepict@height)
\put(0,0){\psfig{figure=#1,width=#2,height=#3,clip=}}
\SetScale{0.283466457}
\SetWidth{1.763889}
{#4}
\end{picture}}
}
\newcounter{pict@widthfst}
\newcounter{pict@widthscd}
\newcounter{pict@widthtot}
\newcommand{\psfigaddtwo}[7]{%
\setcounter{pict@widthfst}{1*\ratio{#2+\pict@scale/2}{\pict@scale}}
\setcounter{pict@widthscd}{1*\ratio{#2+#4+\pict@scale/2}{\pict@scale}}
\setcounter{pict@widthtot}{1*\ratio{#2+#4+#6+\pict@scale/2}{\pict@scale}}
\setcounter{pict@height}{1*\ratio{#3+\pict@scale/2}{\pict@scale}}
\setlength{\unitlength}{\pict@scale}
\hbox{\hspace{-\fill}\begin{picture}(\thepict@widthtot,\thepict@height)
\put(0,0){\psfig{figure=#1,width=#2,height=#3,clip=}}
\put(\thepict@widthscd,0){\psfig{figure=#5,width=#6,height=#3,clip=}}
\SetScale{0.283466457}
\SetWidth{1.763889}
{#7}
\end{picture}}
}
\newcommand{\psfigror}[4]{%
\setcounter{pict@width}{1*\ratio{#2+\pict@scale/2}{\pict@scale}}
\setcounter{pict@height}{1*\ratio{#3+\pict@scale/2}{\pict@scale}}
\setlength{\unitlength}{\pict@scale}
\hbox{\begin{picture}(\thepict@width,\thepict@height)
\put(0,\thepict@height){\psfig{figure=#1,width=#3,height=#2,clip=,angle=270}}
\SetScale{0.283466457}
\SetWidth{1.763889}
{#4}
\end{picture}}
}
\newcommand{\psfigrol}[4]{%
\setcounter{pict@width}{1*\ratio{#2+\pict@scale/2}{\pict@scale}}
\setcounter{pict@height}{1*\ratio{#3+\pict@scale/2}{\pict@scale}}
\setlength{\unitlength}{\pict@scale}
\hbox{\begin{picture}(\thepict@width,\thepict@height)
\put(0,0){\psfig{figure=#1,width=#3,height=#2,clip=,angle=90}}
\SetScale{0.283466457}
\SetWidth{1.763889}
{#4}
\end{picture}}
}
\catcode`\@=12 

\begin{document}
\title{Status of the ANTARES Project}
\author{U.~F.~Katz,%
\thanks{Supported by the German Federal Ministry for Education and Research, 
        BMBF, grant no.~05CN2WE1/2} %
for the ANTARES Collaboration}
\institute{\selectlanguage{german}
           Physikalisches Institut, Friedrich-Alexander-Universit"at Erlangen-N"urnberg, 
           Erwin-Rommel-Str.~1, D--91058 Erlangen,\newline
           \email{katz@physik.uni-erlangen.de}}
\date{\strut}
%
\abstract{
The ANTARES collaboration is constructing a neutrino telescope in the
Mediterranean Sea at a depth of 2400 metres, about 40 kilometres off the French
coast near Toulon.  The detector will consist of 12 vertical strings anchored at
the sea bottom, each supporting 25 triplets of optical modules equipped with
photomultipliers, yielding sensitivity to neutrinos with energies above some
$10\gev$. The effective detector area is roughly $0.1\km^2$ for neutrino
energies exceeding $10\tev$.  The measurement of the \v Cerenkov light emitted
by muons produced in muon-neutrino charged-current interactions in water and
under-sea rock will permit the reconstruction of the neutrino direction with an
accuracy of better than $0.3^\circ$ at high energies. ANTARES will complement
the field of view of neutrino telescopes at the South Pole in the low-background
searches for point-sources of high-energy cosmic neutrinos and will also be
sensitive to neutrinos produced by WIMP annihilation in the Sun or the Galactic
centre.
\PACS{
      {95.55.Vj}{Neutrino, muon, pion, and other elementary particle detectors;
                 cosmic ray detectors}
      \and
      {95.35.+d}{Dark matter (stellar, interstellar, galactic, and cosmological)}
      \and
      {95.30.-k}{Fundamental aspects of astrophysics}
     } 
} 
\maketitle
\section{Introduction}
\label{sec-intr}
Due to their weak interactions with matter and radiation, neutrinos are ideal
messengers for the observation of distant astrophysical objects and processes in
environments that are opaque to photons. However, the tiny neutrino cross
sections at the same time require the instrumentation of huge target masses for
neutrino detection, suggesting the use of naturally abundant detection
materials, such as water or ice. Several such projects are currently operational
\cite{astro-ph-0306536,pnl:106:21} or in preparation
\cite{astro-ph-9907432,npps:100:344,nim:a502:150,npps:118:388}.

The ANTARES Collaboration, comprising particle physics, astronomy and sea
science institutes from 7~European countries, is constructing a neutrino
telescope about 40~kilometres off the French Mediterranean coast, at a depth of
2400~metres. First components have been installed, and prototype detector lines
have been deployed and operated between Dec.~2002 and July~2003.  The full
detector will be completed in 2006.

\section{Detector Design and Physics Objectives}
\label{sec-desi}
\begin{figure}[th]
\psfig{file=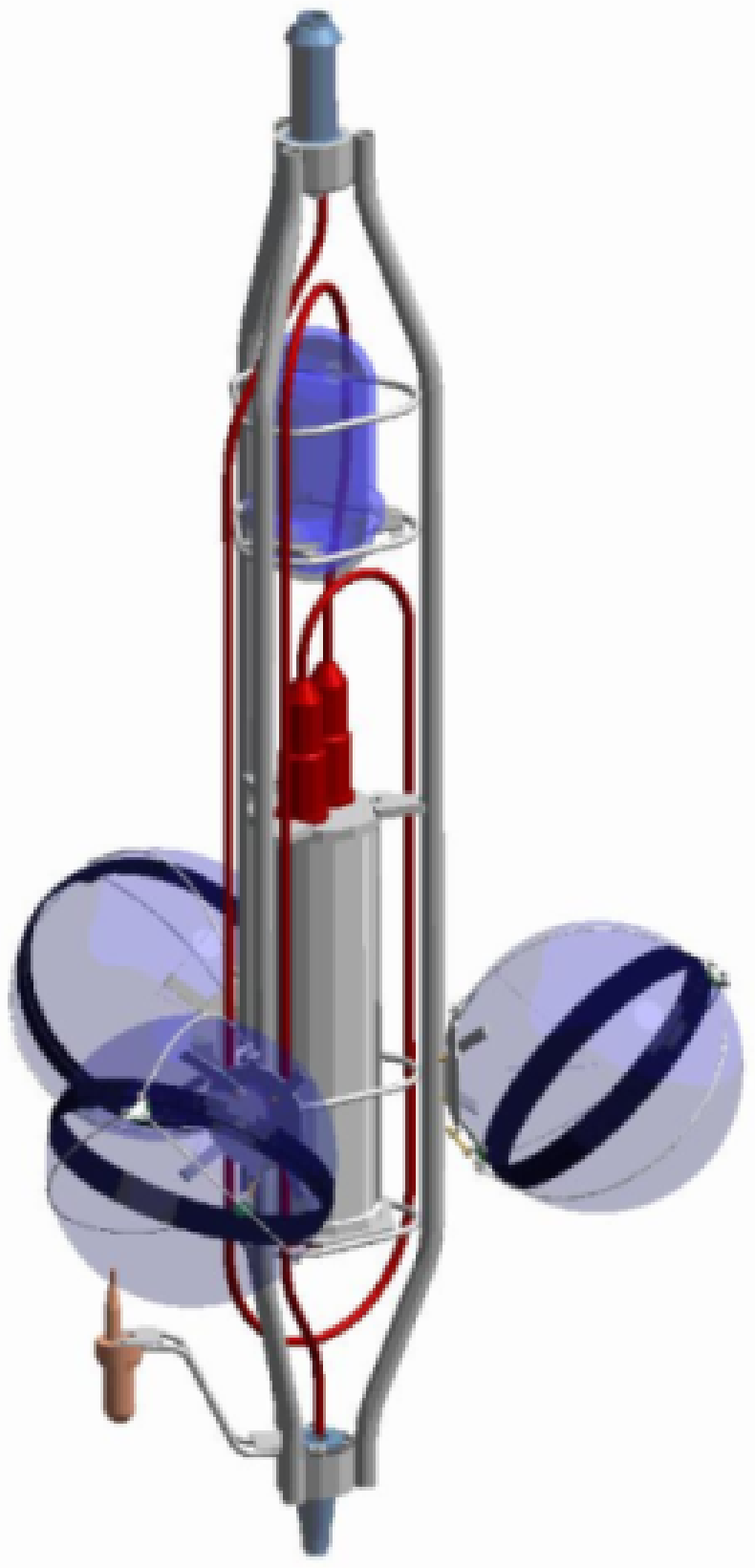,width=4.cm}
\vspace*{-75.mm}

\strut\hspace*{4.0cm}\psfig{file=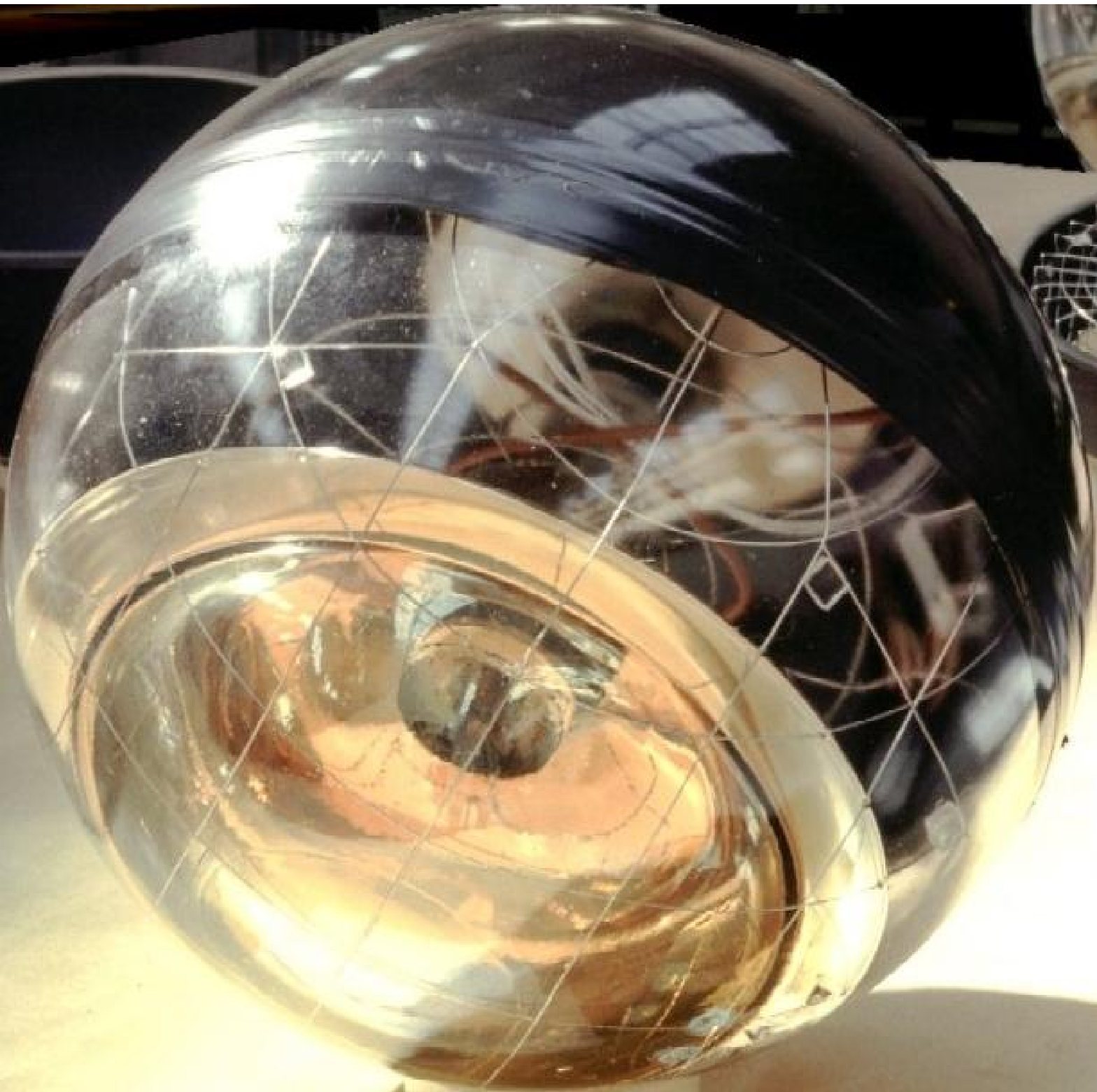,width=4.7cm}
\vspace*{23.mm}

\caption{Left: A storey of a string with the three OMs supported by a titanium
         frame. The cylindrical container for the electronic components (bottom)
         and an optical beacon (top) are located inside the frame. Right: An
         optical module. The metal grid inside the sphere is a mu metal cage
         shielding the PMT against the Earth magnetic field.}
\label{fig-om}
\end{figure}

The ANTARES neutrino telescope \cite{astro-ph-9907432} will detect the \v
Cerenkov light emitted by secondary particles produced in neutrino reactions in
sea water or in the rock below the sea bed. The detector is optimised for
charged-current reactions of muon-neutrinos (yielding a high-momentum muon), but
will also be sensitive to other neutrino flavours and to neutral-current
reactions. It will consist of 12~lines ({\it ``strings''}) that are anchored to
the sea bed at distances of about $70\met$ from each other and kept vertical by
buoys. Each string is equipped with 75 optical modules (OMs) \cite{nim:a484:369}
arranged in triplets ({\it storeys}, see Fig.~\ref{fig-om}) subtended by
titanium frames that also support water-tight titanium containers for the
electronic components. The OMs are glass spheres housing one 10-inch photo
multiplier tube (PMT) each, directed at an angle of $45^\circ$ towards the sea
bed. The storeys are spaced at a vertical distance of $14.5\met$ and are
interconnected with an electro-optical-mechanical cable supplying the electrical
power and the control signals and transferring the data to the string bottom.
Submersible-deployed electro-optical link cables connect the strings to the {\it
junction box (JB)}, which acts as a fan-out between the main electro-optical
cable to shore and the strings. Each string carries optical beacons for timing
calibration and acoustic transponders used for position measurements. The
detector will be complemented by an {\it instrumentation line} supporting
devices for measurements of environmental parameters and tools used by other
scientific communities, such as e.g.\ an seismometer.

The PMTs detect photons with a quantum efficiency above $20\%$ for the relevant
wave lengths between $330\nm$ and $460\nm$. The time resolution is limited by
the transit time spread of about $2.7\ns$ (FWHM). The PMT signals are processed
with custom-designed {\it Analogue Ring Sampler} ASIC chips that measure the
arrival time and charge for signals up to about 10~photo electrons and perform
wave form digitisation for larger pulses. All signals above an adjustable
threshold (usually corresponding to 0.3 photo electrons) are sent to shore,
where an online filter running on a PC farm selects event candidates and reduces
the data volume recorded on tape to about $1\,\text{MB}/\Scnd$.

Muons from muon-neutrino charged-current reactions are finally identified and
reconstructed by offline algorithms.  From the arrival times of the photons at
the PMTs and the OM positions (known to about $5\cm$), the trajectory of muons
can be determined with a pointing precision of about $0.2\rnge0.3^\circ$, which
is the dominant contribution to the experimental uncertainty on the neutrino
direction for neutrino energies $E_\nu\gtrsim10\tev$. The muon energy, $E_\mu$,
is determined from the muon range at small energies and from the \v Cerenkov
intensity due to radiative energy losses at high energies. In the latter case,
the RMS resolution on $\log E_\mu$ is $0.2\rnge0.3$.

\begin{figure}[th]
\begin{center}
\psfig{file=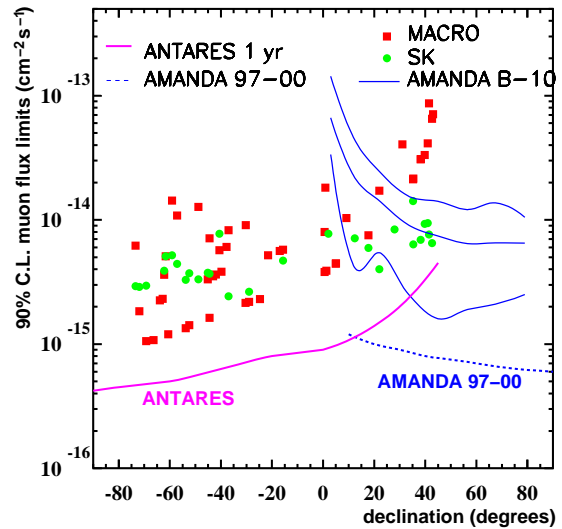,width=0.4\textwidth,clip=}
\end{center}
\caption{ANTARES sensitivity to the neutrino-induced muon flux from
         astrophysical point sources, compared to upper limits from
         other experiments. The simulation is
         based on 1~year of ANTARES operation in the final detector
         configuration.}
\label{fig-ps}
\end{figure}

Detailed simulation studies have been performed to assess the physics
sensitivity of ANTARES. After 3~years of operation, the ANTARES data will
challenge predicted upper limits for diffuse neutrino fluxes
\cite{hep-ex-0308074} and will be sensitive to point source intensities
indicated by different models. In Fig.~\ref{fig-ps}, the expected ANTARES
sensitivity to muon flux induced by neutrinos from point sources is compared to
upper limits from other experiments. Note in particular the complementary sky
coverage of AMANDA and ANTARES. The search for neutrinos from gravitational
centres, such as the Sun or the galactic centre, yields sensitivity to WIMP
annihilation and thus complements direct-search experiments
\cite{misc:icrc03:an:1}.

\section{The Preparatory Phase}
\label{sec-prep}
Since the ANTARES neutrino telescope has to operate in the uncontrollable
deep-sea environment of the Mediterranean, the environmental conditions are of
prime importance to the site choice and to the detector design. The ANTARES
Collaborations has performed a series of more than 40~deployments of explorative
devices and prototype modules for site investigations and technical studies. At
the ANTARES site, direction and speed of the deep-sea currents, the water
transparency, the background rates caused by \nucl{K}{40} decays and
bioluminescence, the sedimentation, the bio-fouling of the OM surfaces and
various other parameters have been repeatedly measured and have been found
consistent with the requirements for the detector operation
\cite{misc:icrc:1999:492,*app:13:127,*app:19:253}.

In 1999, a {\it demonstrator string} equipped with 7~PMTs, full readout
electronics, slow-control devices and an acoustic positioning system was
deployed at a depth of $1100\met$, connected to an existing telecommunication
cable to shore and operated for 8~months. The position and timing calibration
have been demonstrated to work as expected, and about $5\times10^4$~7-fold
coincidences from atmospheric muons have been recorded.

\section{Detector Status}
\label{sec-test}
\begin{figure*}[t]
\sidecaption
\psfig{file=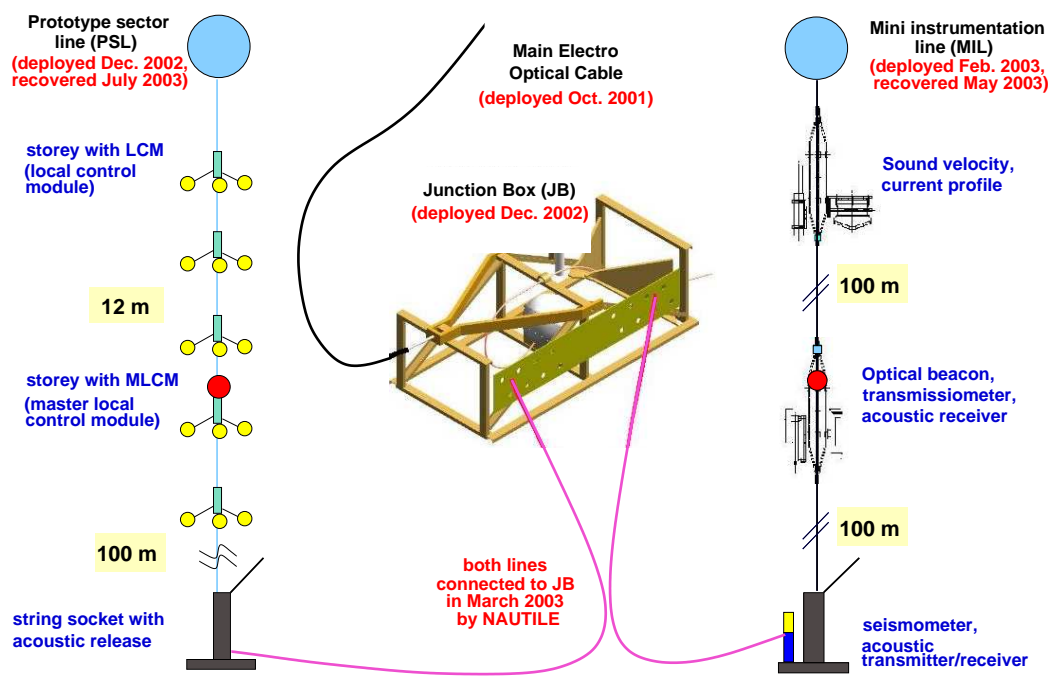,width=0.77\textwidth,clip=}
\caption{\protect\raggedright
         Components of the ANTARES neutrino telescope and prototype strings
         deployed by summer~2003.}
\label{fig-pr}
\end{figure*}

The current status of the ANTARES detector is indicated in Fig.~\ref{fig-pr}.
The first component of the final detector configuration, the main
electro-optical cable, has been deployed in Oct.~2001 and connected to the shore
station. In Dec.~2002, the end of the cable was recovered from the sea-bed, the
JB connected to it and subsequently deployed. Communication with the JB slow
control is being maintained since then.

Two prototype strings, an optical line with 5~storeys ({\it prototype sector
line, PSL}) and the {\it mini instrumentation line (MIL}), were deployed in
Dec.~2002 and Feb.~2003, respectively, and connected to the JB in an undersea
operation by the manned submersible {\it Nautile} in March~2003. Communication
with both lines was established immediately after connection. The systems were
found to be functional with the exceptions detailed below. A large bulk of data
has been acquired and is currently analysed. The lines have been recovered in
May (MIL) and July~2003 (PSL).

Two problems occured in the prototype tests: In the MIL, a water leak developed
in one of the electronic containers, leading to a short circuit in the power
supply that made further operation impossible. After recovery of the MIL in
May~2003 a faulty supplier specification for the installation of a connector was
identified as reason for the leak; the design has been modified to exclude this
problem in the future. The second problem was that in both lines the clock
signal, sent from shore to the off-shore electronic modules to synchronise the
readout, reached the bottom string socket (BSS) but not the detector. In the MIL,
the clock failure was due to a damaged glass fibre in the cable between BSS and
first storey, caused by the supplier's use of an unsuited material for the fibre
coating. In the PSL, the corresponding investigation is still on-going.

Due to the absence of the clock signal, no data with timing information at
nanosecond precision could be taken. Nevertheless, the long-term operation of
the PSL over more than 3~months yielded a wealth of information, both on the
functionality of the detector and the environmental conditions. In particular,
the rate of signals above threshold was monitored continuously for each OM. It
was found that the rates exhibit strong temporal variations that are attributed
to bioluminescent organisms. A continuous rate, varying between about $50\kHz$
and $250\kHz$ per OM, is accompanied by short light bursts that cover between
less than $1\%$ and more than $30\%$ of the overall time. Also monitored were
the heading and tilt of the PSL storeys. It was found that they move almost
synchronously, i.e.\ the PSL behaves as a pseudo-rigid body in the water
current.  Correlations of the background rates with the movement of the PSL and
hence with the sea currents have been observed. Detailed investigations of the
on-line filter requirements imposed by high rates and of relations between water
currents, string movements and bioluminescence are under way.

\section{Conclusion and Outlook}
\label{sec-conc}
With the installation of the main electro-optical cable and the junction box,
the ANTARES project has entered the construction phase. Prototype detector
strings have been successfully deployed and operated, verifying the detector
design and functionality and yielding a vast amount of environmental
data. Failures that occured in a connector to one of the electronics containers
and in the transmission of the clock signal will be avoided in the future by
implementing modest design modifications. For an ultimate verification of all
elements of the detector design, it is forseen to deploy a new instrumentation
line combining instrumentation and optical modules in mid-2004.  The first final
detector string will be installed by the end of 2004. First physics data are
expected by 2005, the completion of the detector is scheduled for 2006.

\medskip\noindent
{\bf Acknowledgements:} I would like to thank the organisers of the {\sc
HEP2003} conference for a very inspiring week in Aachen and the conveners of
the parallel sessions for their help and support in adjusting the schedule to
the needs of the speakers.


{\raggedright
\bibliographystyle{hep03}
\bibliography{hep03}
}

\end{document}